\documentclass[%
 prl,
 aps
 amsmath,
 amssymb,
reprint,%
superscriptaddress,
showpacs
]{revtex4-1}

\usepackage{graphicx}
\usepackage{dcolumn}
\usepackage{bm}
\usepackage{amsmath} 

\begin{document}


\title[Two-Way Optical Frequency Comparisons Over 100\,km Telecommunication Network Fibers]{Two-Way Optical Frequency Comparisons\\ Over 100\,km Telecommunication Network Fibers}

\author{Anthony Bercy}
\affiliation{
Laboratoire de Physique des Lasers, Universit\'{e} Paris 13, Sorbonne Paris Cit\'{e}, CNRS, \\99 Avenue Jean-Baptiste Cl\'{e}ment, 93430 Villetaneuse, France
}%
\affiliation{ 
Laboratoire National de M\'{e}trologie et d'Essais-Syst\`{e}mes de R\'{e}f\'{e}rence Temps-Espace, UMR 8630 \\Observatoire de Paris, CNRS, UPMC, 61 Avenue de l'Observatoire, 75014 Paris, France}
\author{Fabio Stefani}
\affiliation{ 
Laboratoire National de M\'{e}trologie et d'Essais-Syst\`{e}mes de R\'{e}f\'{e}rence Temps-Espace, UMR 8630 \\Observatoire de Paris, CNRS, UPMC, 61 Avenue de l'Observatoire, 75014 Paris, France}
\author{Olivier Lopez}
\affiliation{
Laboratoire de Physique des Lasers, Universit\'{e} Paris 13, Sorbonne Paris Cit\'{e}, CNRS, \\99 Avenue Jean-Baptiste Cl\'{e}ment, 93430 Villetaneuse, France
}%
\author{Christian Chardonnet}
\affiliation{
Laboratoire de Physique des Lasers, Universit\'{e} Paris 13, Sorbonne Paris Cit\'{e}, CNRS, \\99 Avenue Jean-Baptiste Cl\'{e}ment, 93430 Villetaneuse, France
}%
\author{Paul-Eric Pottie}
\email{paul-eric.pottie@obspm.fr}
\affiliation{ 
Laboratoire National de M\'{e}trologie et d'Essais-Syst\`{e}mes de R\'{e}f\'{e}rence Temps-Espace, UMR 8630 \\Observatoire de Paris, CNRS, UPMC, 61 Avenue de l'Observatoire, 75014 Paris, France}
\author{Anne Amy-Klein}
\affiliation{
Laboratoire de Physique des Lasers, Universit\'{e} Paris 13, Sorbonne Paris Cit\'{e}, CNRS, \\99 Avenue Jean-Baptiste Cl\'{e}ment, 93430 Villetaneuse, France
}%
\date{\today}

\begin{abstract}
By using two-way frequency transfer, we demonstrate ultra-high resolution comparison of optical frequencies over a telecommunication fiber link of 100 km operating simultaneously digital data transfer. We first propose and experiment a bi-directional scheme using a single fiber. We show that the relative stability at 1\, s integration time is $7\cdot10^{-18}$ and scales down to $5\cdot10^{-21}$. The same level of performance is reached when an optical link is implemented with an active compensation of the fiber noise. We also implement 
a real-time two-way frequency comparison over a uni-directional telecommunication network using a pair of parallel fibers. The relative frequency stability  is $10^{-15}$ at 1\,s integration time and reaches $2\cdot10^{-17}$ at 40\,000\,s. The fractional uncertainty of the frequency comparisons was evaluated for the best case to $2\cdot 10^{-20}$. These results open the way to accurate and high resolution frequency comparison of optical clocks over intercontinental fiber networks.%
\end{abstract}


\maketitle
 
High resolution time and frequency transfer between remote locations are of major interest for many applications, such as tests of general relativity and temporal variation of fundamental constants, future redefinition of the second, relativistic geodesy and navigation (see\,\cite{Giorgetta:2013} and references herein). It is usually performed through satellite-based time and frequency transfer but with performance now insufficient for state-of-the-art optical clocks and laser oscillators\,\cite{Bloom:2014, Kessler:2012,Fujieda:2014}. As a very promising alternative, optical fiber links are intensively studied by several groups for frequency transfer for a decade\,\cite{Newburry:2007,Predehl:2012,Lopez:2012}. They demonstrate impressive results far beyond the GPS capabilities on distances up to 1840\,km with bi-directional dedicated fibers\,\cite{Droste:2013}. Our groups extended the technique of optical link to active telecommunication fiber networks by inserting  Optical Add \& Drop multiplexers (OADM) in every amplification sites and network nodes. We enable  bi-directional frequency transfer on one dedicated channel of 100\,GHz, in parallel with uni-directional data traffic over all the other channels\,\cite{Kefelian:2009}. This is a very efficient technique for ultra-stable time and frequency transfer on a continental scale\,\cite{Lopez:2013}. Our approach gives the possibility to route a frequency standard signal on a broad fiber network, aiming at frequency standard dissemination to a wide number of laboratories, for high-resolution spectroscopy, remote laser stabilization, and any high precision measurements.%

If one focus on optical frequency comparisons, and let the frequency transfer aside, the set up can be drastically simplified with two-way method\,\cite{Hanson:1989}. At each end of the link, a laser is sent to the other end and one detects the frequency difference between the local laser and the remote laser. Assuming that the propagation frequency noise is equal for the two directions of propagation, one can efficiently reject the propagation contributions by post processing the data, simply subtracting and dividing by two the two data sets recorded at each end. Two-way frequency comparison was recently demonstrated  for two laboratories in the same place over a 47\,km loop in an urban link\,\cite{Calosso:2014}. We demonstrate here two schemes of two-way frequency comparisons over 100\,km urban link that can be practically implemented with two distant laboratories. The first uses a single fiber through which the light is propagated in both directions (from here referred to as two-way bi-directional or 2way-B). It exhibits limitation similar to that of an actively noise compensated link for frequency comparisons. In addition, we 
realize a second scheme which uses two parallel fibers, each fiber transmitting the light in a single direction (referred to as two-way uni-directional or 2way-U)\,\cite{Williams:2008}. This uni-directional scheme opens the way to frequency comparisons over telecommunication network with minimal modification of the network backbone. 

The paper is organized as follows\,: we describe first the fiber network infrastructure and the two schemes of the two-way frequency comparisons we have implemented. Second we present the experimental results and the extrapolation to long haul terrestrial and submarine links of uni-directionnal two-way frequency comparisons. 

In order to demonstrate in real conditions two-way frequency comparison and to asset its performances, we used a pair of optical fibers connected as two parallel loops of 100\,km in Paris area. The loop starts and ends at Laboratoire de Physique des Lasers and is constituted of six fibers spans. In each span, we access two fibers placed in the same cable.  Two non-consecutive spans of 10\,km and 8\,km are active telecommunication fibers from French Research Network (Renater), simultaneously used for data traffic. 
Eight OADMs are used to insert and to extract the ultra-stable signal at 1.5\,$\mu$m into these spans. The four others spans, that are 82\,km long, are dedicated fibers. We installed halfway one Erbium-Doped Fiber Amplifiers (EDFA) on each fiber with 20\,dB gain to compensate partly the 45\,dB losses. 
\begin{figure}
\includegraphics[width=7cm]{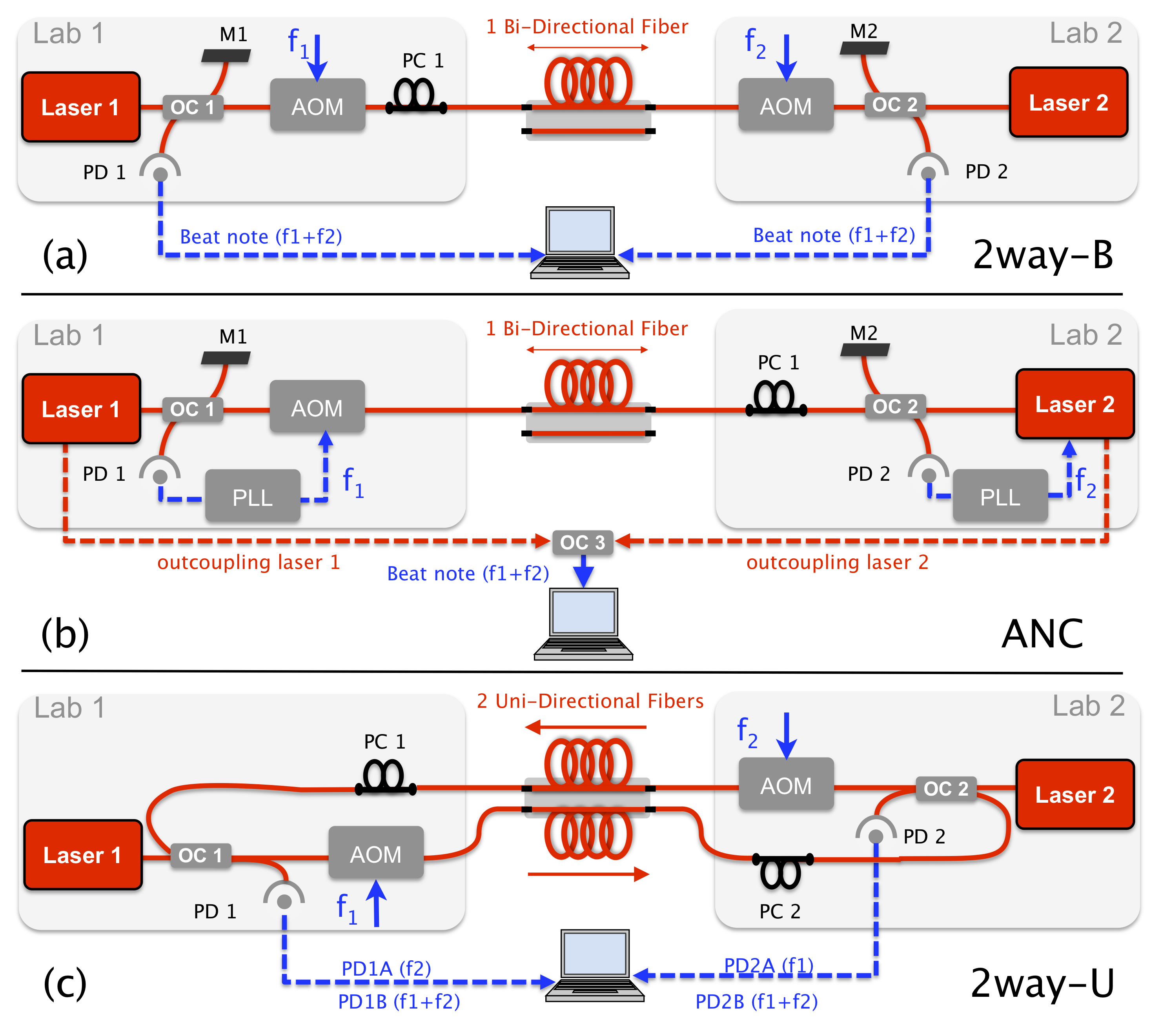}
\caption{\label{Fig:sketch}
Experimental set ups for frequency comparison: a) two-way bi-directional  b) actively noise compensated link c) two-way uni-directional. AOM : Acousto-Optic Modulator. PC : Polarization Controller. FM : Faraday Mirror. OC : Optical Coupler. PD : Photo-Diode PLL : Phase Lock Loop. }
\end{figure}

First we tested a two-way bi-directional configuration (see Fig. \ref{Fig:sketch}a). The frequency of the two lasers located at each end are shifted with two Acousto-Optics Modulators (AOM) at frequencies $f_1$ and $f_2$ in order to distinguish the useful signals from parasitic back-reflections. At each end, two beat notes at frequency $f_1+f_2$ between the local and remote lasers are obtained, using two independent Michelson-type interferometers. One polarization controller is used to optimize the beat notes. We carefully designed the detection interferometers so that the temperature-induced noise at low frequencies is below the measured long-term noise floor of the system (a few fs on a typical day/night cycle)\,\cite{Stefani:2014}. 
After detection, amplification and filtering, the beat notes are simultaneously recorded with a gate time of 1\,s, with two dead-time free frequency counters (Kramer + Klische FXE) operated in $\Pi$-type and $\Lambda$-type. These two sets of one-way data are then subtracted and divided by two to obtain the optical frequency comparison. With a fiber loop, the two ends  are co-located in the same laboratory. We asset the ultimate performances of this set-up by injecting both ends with a single fiber laser. The two interferometers are hosted in a single, thermally controlled box, so that most of the interferometer noise is rejected\,\cite{Stefani:2014}. 

Following the general approach of Newbury \,\cite{Newburry:2007}, and using the Fourier transform of the autocorrelation function as described in\,\cite{Bercy:2014}, we can derive the expected high-frequency noise (typically $f>$1\,Hz) of a bi-directional two-way set-up. It has a similar origin to the delay-unsuppressed noise in a noise compensated link. On Fig.\,\ref{Fig:sketch}, the optical signals propagating from Lab\,1 to Lab\,2 and from Lab\,2 to Lab\,1 are denoted with underscript {\it 12} and {\it 21}. 
The total phase perturbation  $\Phi_{12}$ (respectively $\Phi_{21}$) reads :
$\Phi_{12} (t)= \int_0^{L} \delta \Phi \left(z, t-\frac{L-z}{v}\right)\,dz\nonumber$ 
and $\Phi_{21}(t) = \int_0^{L} \delta \Phi \left(z, t-z/v\right)\,dz\nonumber$,
where $v$ is the celerity of light in the fiber, $L$ is the length of the loop,  and $\delta \varphi$ is the phase perturbation per unit of length at coordinate $z$ at time $t$.
Assuming that the Fourier frequency times the round-trip time is small, the two-way phase $\Phi_{\textrm{tw}}=1/2(\Phi_{12}(t)-\Phi_{21}(t))$ is at first order\,:
\begin{eqnarray}
\label{Eq:PhiTW}
\Phi_{\textrm{tw}}(t) =\frac{1}{2}\int_0^{L} \delta \varphi '(z, t) \cdot \left(\frac{2z-L}{v} \right)\,dz
\end{eqnarray}
Calculating the autocorrelation function, $R_{\textrm{tw}}(\tau)=\overline{\Phi_{\textrm{tw}}(t) \Phi_{\textrm{tw}}(t+\tau) }$ and assuming that the fiber noise is not correlated in position and independent of z, one derives\,:
\begin{equation}
\widetilde{ \Phi_{\textrm{tw}}}(j\omega)=\mathcal{F}( R_{\textrm{tw}}(\tau))=\widetilde{\Phi_{12}}(j\omega)\cdot \frac{(2\pi f \cdot L/v)^{2}}{12}
\end{equation}
This formula is very similar to that obtained for an optical link with active noise compensation, but with a factor $1/12$ instead of $1/3$, since the two-way phase is half of the difference between the two one-way phase signals\,\cite{Newburry:2007,Calosso:2014}

In order to prove the effectiveness of this statement, we set up on the same 100\,km fiber loop an active noise compensated link, (see Fig. \ref{Fig:sketch}b) referred to as {\it ANC} later on\,\cite{Ma:1994}. We recorded the  beat note between the two ends of the link at the same frequency $f_1+f_2$. A noticeable change is that the fiber laser is then phase locked to an ultra-stable laser, itself locked to an ultra stable cavity, transferred from SYRTE to LPL on a 43\,km long dedicated fiber\,\cite{Jiang:2008}. 

The bi-directional two-way technique described above uses one fiber with bi-directional propagation whereas the telecommunication network operates uni-directional propagation. Aiming at a simplified optical frequency transfer and at developing a broader fiber network, and following a first proposal reported in\,\cite{Williams:2008}, we are 
implementing here another technique with two-fibers and uni-directional propagation. The two-way uni-directional set up uses two fibers from the same cable. This scheme complies with the network backbone, except for  switches and routers. The set-up is sketched on Fig. \ref{Fig:sketch}c. The left to right and right to left counter-propagating optical signals are separated into two uni-directional fiber paths. One additional optical amplifier was added on the second fiber to compensate extra-losses we observed on it. A second polarization controller was also inserted to optimize the beat note. Otherwise the set-up is identical to the 2way-B set-up.

To point out the particularities of this approach, let us consider a simplified, steady state model of the oscillators and link noise. Considering that on the left hand side we have an AOM driven at a frequency $f_1$ on one fiber, and at the right hand side a second AOM at a frequency $f_2$ on the second fiber, one can write the beat notes detected by the photodiodes PD1 and PD2 as follows\,:
\begin{eqnarray}
   \left\{
    \begin{aligned}
 	\textrm{PD1}_A  &=(\Phi_2+\Phi_{21})-\Phi_1 &\textrm{at frequency }f_2 \\
 	\textrm{PD1}_B &=(\Phi_{12}+\Phi_{21}) &\textrm{at frequency }f_1+f_2  \\
 	\textrm{PD2}_A &=(\Phi_1+\Phi_{12})-\Phi_2 &\textrm{at frequency }f_1 \\
 	\textrm{PD2}_B &=(\Phi_{12}+\Phi_{21}) &\textrm{at frequency }f_1+f_2 \\
    \end{aligned}
    \right.
 \label{Eq:set2wayMonoDir}
\end{eqnarray}
where $\Phi_1$ and $\Phi_2$ are the phase noise associated to Laser\,1 and Laser\,2 and  $\Phi_{12}$ and $\Phi_{21}$ the noise added by the two fibers respectively (independent of the lasers at first order). The signals labeled as {\it A} are derived from the beat note between the  local and remote lasers; the signals labeled as {\it B} are derived from the beat note between a local laser with itself after a round trip in the fiber loop. It can be easily demonstrated that, for any realistic amount of fiber attenuation, the successive loops of light signals are not perturbing the considered measurements.

Assuming that the phase noise between both fibers is partly correlated, one expects a partial noise cancellation for a frequency comparison. Combining the signals labelled as {\it A} in Eq.\,\ref{Eq:set2wayMonoDir} in post-processing gives\,:
\begin{equation}\label{Eq:2way-u}
(-\textrm{PD1}_A +\textrm{PD2}_A)/2 = (\Phi_1-\Phi_2)+(\Phi_{12}-\Phi_{21})/2
\end{equation}
Under the assumption that $\Phi_{12} =  \Phi_{21}$,  one finds the standard two-way noise rejection, as  the fiber noise cancels out and only the laser's phase difference remains. In addition, the same results can be obtained by combining the two beat notes falling on the same photodiode. For instance at the end\,1 on PD\,1 one has\,:
\begin{equation}\label{Eq:2way-u-local}
 -\textrm{PD1}_A +\textrm{PD1}_B /2 = (\Phi_1-\Phi_2)+(\Phi_{12}-\Phi_{21})/2
\end{equation}
and similarly at end\,2 on PD\,2. This approach, that we called "local two-way", allows to process the noise rejection at both distant laboratory, relying only on data acquired locally. It avoids the necessity to exchange and synchronize data between distant sites. The main limiting factor of the 2way-U arises from the uncorrelated fiber's noise, that causes the basic assumption of the two-way method to be partially violated.

The performance of this set-up was addressed by using the same fiber laser at both ends. The beat notes at frequency $f_1$ or $f_2$ are amplified and filtered before being recorded with counters. Those at frequency $f_1+f_2$ are more attenuated because of a double circulation in the loop. They are amplified, filtered, and then tracked with a tracking oscillator in a bandwidth of 100\,kHz.
The combination of recorded data from beat note at $f_1$ and $f_2$ from Eq.\,\ref{Eq:2way-u}, gives the two-way frequency comparison. The real-time combination of recorded data from beat note $f_2$ and $f_1+f_2$ from Eq.\,\ref{Eq:2way-u-local} (and resp. $f_1$ and $f_1+f_2$)  gives what we called a {\it local} frequency comparison. 

\begin{figure}[t]
\includegraphics[width=7cm]{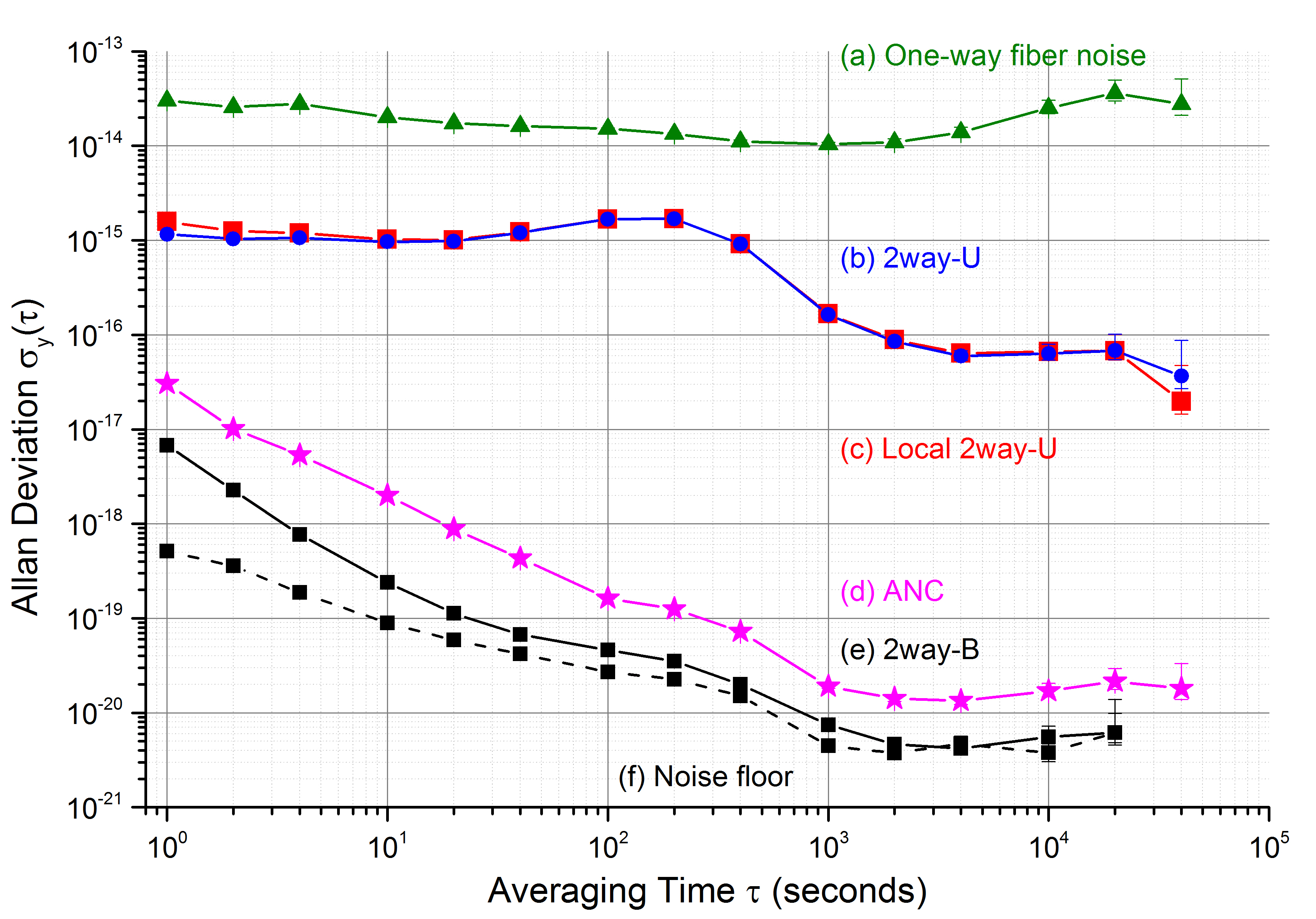}
\caption{\label{Fig:MDEV}
Fractional frequency instability derived from data recorded with $\Lambda$-counter and expressed as the modified Allan deviation for (a) One-way fiber noise (b) Two-way uni-directional as reconstructed from data recorded at the 2 ends (c) Local two-way uni-directional  (d) Active noise compensated link (e) Two-way bi-directional. (f) Two-way bi-directional noise floor. The other noise floors are similar and are not shown for sake of clarity.}
\end{figure}
 
We plot on Fig. \ref{Fig:MDEV} the instability of the two-way frequency comparison using either the uni-directional or the bi-directional set-up, both expressed as the modified Allan deviation (MDEV)\,\cite{McFerran:2007}. Note that for both 2way-U and 2way-B configurations, we did not suppress any point from the data sets. The two-way bi-directional stability is $7\cdot 10^{-18}$ at 1\,s and reaches $5\cdot 10^{-21}$ at  4\,000\,s. The 2way-B noise floor is below the reported stabilities at short-term, but limits the the 2way-B stability for $\tau>$100\,s. We checked that, at long term, it was not induced by the temperature sensitivity of the interferometer\,\cite{Stefani:2014}. The 2way-B relative stability is about four times below the one of the ANC set-up, which is interestingly more than expected. Moreover the local two-way uni-directional stability is as low as $10^{-15}$ at 1\,s integration time and reaches $2\cdot 10^{-17}$ at 40\,000\,s. The remote 
2way-U has almost the same performance. This stability is limited by the uncommon fiber noise, thus scaling as $\sqrt L$. It is demonstrating its excellent capabilities for frequency comparison over Telecommunication network under operation, as long as the switches and routers are bypassed. This level of performance is indeed already far beyond the most advanced GPS and Two-Way Carrier Phase capabilities\,\cite{Fujieda:2014}. 

\begin{figure}[t]
\includegraphics[width=8cm]{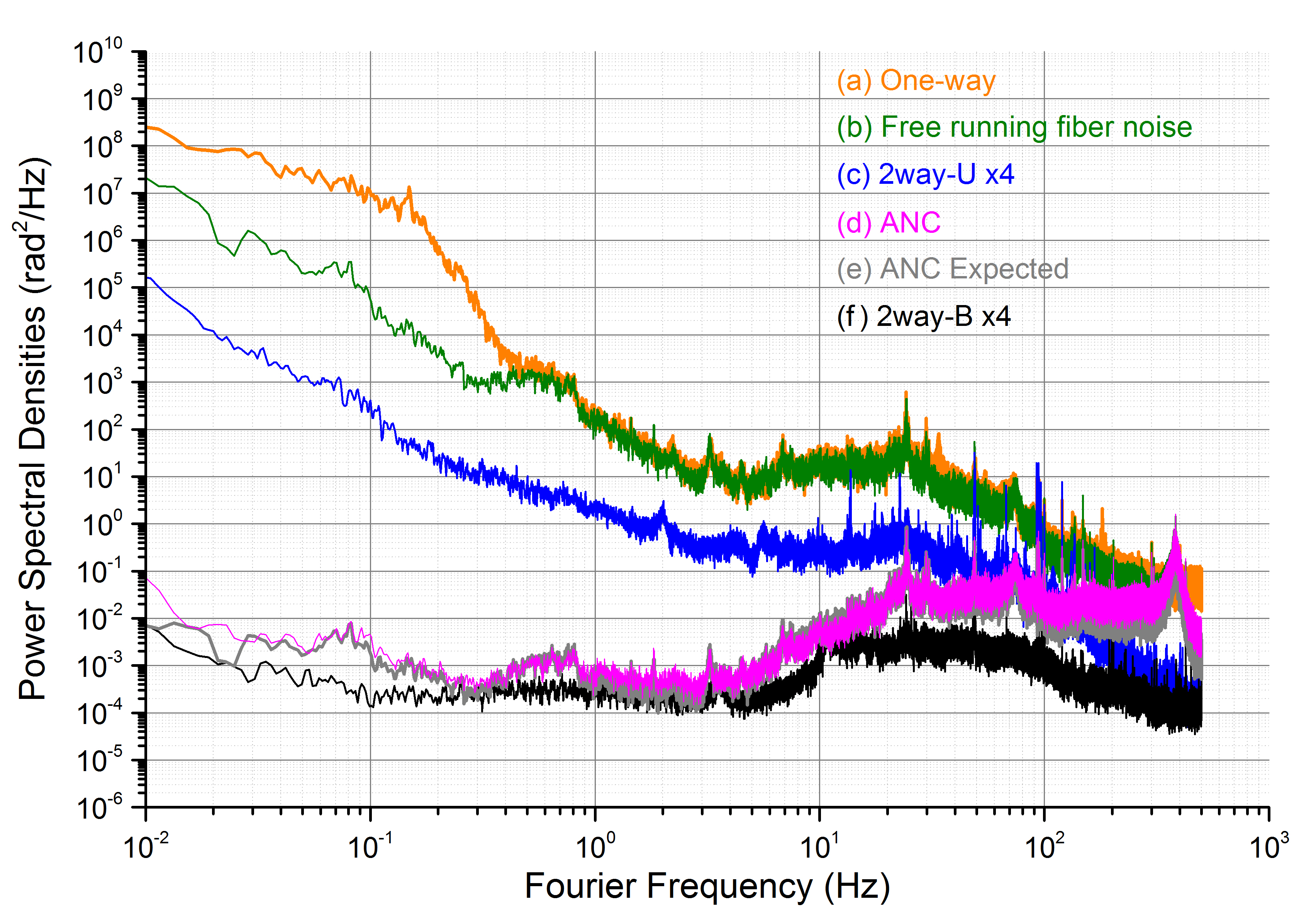}
\caption{\label{Fig:PSD}
Phase noise PSD for the 3 set-ups: (a) One-way, (b) Free running noise for the ANC set up (c) two-way uni-directional, where real-time and post-processed overlapped themselves ($\times 4$), (d) Active noise compensation, (e) ANC expected (f) two-way bi-directional ($\times 4$).}
\end{figure}

To further investigate the noise rejection of the three set-up sketched on Fig.\,\ref{Fig:sketch}, we plot on Fig. \ref{Fig:PSD} their phase noise Power Spectral Densities (PSD). The measurements were done with a K+K frequency counters with a gate time of 1\,ms and $\Pi$-type operation. Frequency data were converted to phase data. The two-way PSD were scaled by a factor 4 ({\it i.e.}+6\,dB) in order to compare them easierly with the ANC PSD. 
The excess of noise PSDs we observe between the one-way and the free-running fiber noise of the ANC set-up for $f <5\cdot 10^{-1}$\,Hz is due to the free running laser we used for the two-way set-ups. This is worth to notice that this common noise source for the two counter propagating signals is well rejected with the two-way set-ups, even for the 2way-U set up that exhibits higher residual phase noise than the 2way-B. The ANC PSD is overlapping with the expected ANC residual noise. The ANC and 2way-B curves shows similar behavior in fair agreement with the theoretical expectations. Interestingly we observe an overcorrection compared with the expectations for the 2way-B. This rejection anomaly shows that the assumptions we made on Eq.\,\ref{Eq:PhiTW}, and that we made on homogeneous noise and uncorrelated noise in position, are violated at some point. Further investigations are needed on the noise correlation properties in such 2-way bi-directional set up. In an urban area network, we are indeed observing greater acoustic noise per unit of length compare to optical links deployed in field.

We now extrapolates our result on 2way-U set up to a 800\,km link, connecting Paris to London for instance. Assuming an homogeneous fiber noise with a deviation scaling as $\sqrt L$, we are expecting a MDEV of $9 \cdot 10^{-17}$ at 30\,000 s integration time. When considering a transatlantic link, one has to take into account that the noise deviation of a submarine link is about 10 times smaller\,\cite{Ebenhag:2011}. We  consider a link constituted of 6\,500\,km submarine link and 1\,000\,km terrestrial link as a typical U.S. to E.U link. We found an expected MDEV of $5\cdot 10^{-15}$ at 1\,s and $1\cdot 10^{-16}$ at 30\,000 s integration time, dominated by the terrestrial noise. We checked that the delay unsuppressed noise that scales as $L^{3/2}$ is not dominating the uni-directional noise. This extrapolation must be confirmed with realistic data on fiber losses and on submarine amplifier's gain. The cumulated spontaneous emission of the optical amplifiers (up to 75) can be the limiting factor of such a method. 

Finally we evaluate the accuracy of the frequency comparison at the remote end. We calculate the mean frequency of the relevant beat notes (for each set up), recorded with 1\,s gate time and $\Pi$-type counter, to avoid overweighting the center parts of the data sets with the triangular weighting of the $\Lambda$-type counters.Then we calculate the mean value of these frequencies and its standard deviation for consecutive segments from 1 to 1\,000\,s\,\cite{Droste:2013}. For the bi-directional set up, with a set of 138\,000 data of 1\,s of Fig.\,\ref{Fig:MDEV}, we obtained a mean offset frequency of $7\cdot10^{-21}$. Using consecutive segments of up to 1\,000 s, the statistical relative uncertainty of the mean value, calculated as the relative standard deviation divided by the length of the consecutive segment, has a constant value of $3\cdot 10^{-21}$ as expected for white phase noise\,\cite{Lee:2010}. For longer segments, this value slightly increases, due to long-term Flicker noise. We finally set a conservative estimate of the statistical fractional uncertainty of the frequency comparison as the long-term overlapping Allan deviation of the data set, which is $2\cdot 10^{-20}$ at 20\,000\,s integration time. For the two-way uni-directional, using 160\,000 data of 1\,s of the remote comparison, we find a relative mean offset frequency of $8\cdot 10^{-18}$. Due to the behavior of the reported MDEV, we set the statistical fractional uncertainty of the frequency comparison with the overlapping Allan deviation, that is $ 6.5\cdot 10^{-17}$ at 40\,000\,s integration time. The frequency comparison shows no deviation  within this uncertainty.

In summary we demonstrate in this article two 
set-ups to compare optical frequencies between two distant laboratories at ultra-high resolution and short averaging time on a 100\,km telecommunication network. The most performant one uses two-way frequency transfer with bi-directional operation in a single fiber. 
We demonstrate accurate and high-resolution frequency comparison with a very simple electronics, with a relative stability of $5\cdot10^{-21}$ at 4\,000\,s integration time in 1\,Hz bandwidth. 
The in-field implementation of two-way bi-directional techniques requires two ultra-stable lasers at each end, and an accurate control of their frequency drifts, which may limit the comparison. This can be done actively by locking the laser frequency to an atomic clock or a Maser, or passively by time stamping the data of the frequency measurements in the 2 laboratories. The uncertainty will be then determined by the linear drift of the lasers, the length of the link and the accuracy of the time stamps.

The second set-up we have presented uses two fibers for each propagation way and comply with uni-directional amplifiers used in telecommunication networks. It reduces considerably the work needed to implement in field a high resolution frequency comparison. We demonstrated a frequency comparison with a relative frequency stability of $2\cdot 10^{-17}$ at 40\,000\,s integration time in 1\,Hz bandwidth. This two-way uni-directional method gives the possibility to perform measurements {\it in situ} and real-time. Despite its lower relative stability, this technique opens the way to intercontinental clocks comparison with fiber links in parallel with data traffic at a level of resolution and accuracy competitive with the most advanced satellite techniques and with much shorter integration time.

\begin{acknowledgments}
The authors thanks Giorgio Santarelli for helpful and stimulating discussions. They also thank Emilie Camisard, Thierry Bono and Patrick Donath at the GIP Renater, and Fran\c cois Biraben, Fran\c cois Nez and Sa\"\i da Guellati-Khelifa at Laboratoire Kastler-Brossel for giving us the opportunity to use the different spans of the fiber's loop.

This work is supported by the European Metrology Research Programme (EMRP) under SIB-02 NEAT-FT. The EMRP is jointly funded by the EMRP participating countries within EURAMET and the European Union. This work is supported by the E.U. under GN3+, the Labex FIRST-TF, the french spatial agency CNES, IFRAF-Conseil R\'{e}gional Ile-de-France and Agence Nationale de la Recherche (ANR BLANC 2011-BS04-009-01).
\end{acknowledgments}

\bibliography{2way}

\providecommand{\noopsort}[1]{}\providecommand{\singleletter}[1]{#1}%
\begin{thebibliography}{20}%
\makeatletter
\providecommand \@ifxundefined [1]{%
 \@ifx{#1\undefined}
}%
\providecommand \@ifnum [1]{%
 \ifnum #1\expandafter \@firstoftwo
 \else \expandafter \@secondoftwo
 \fi
}%
\providecommand \@ifx [1]{%
 \ifx #1\expandafter \@firstoftwo
 \else \expandafter \@secondoftwo
 \fi
}%
\providecommand \natexlab [1]{#1}%
\providecommand \enquote  [1]{``#1''}%
\providecommand \bibnamefont  [1]{#1}%
\providecommand \bibfnamefont [1]{#1}%
\providecommand \citenamefont [1]{#1}%
\providecommand \href@noop [0]{\@secondoftwo}%
\providecommand \href [0]{\begingroup \@sanitize@url \@href}%
\providecommand \@href[1]{\@@startlink{#1}\@@href}%
\providecommand \@@href[1]{\endgroup#1\@@endlink}%
\providecommand \@sanitize@url [0]{\catcode `\\12\catcode `\$12\catcode
  `\&12\catcode `\#12\catcode `\^12\catcode `\_12\catcode `\%12\relax}%
\providecommand \@@startlink[1]{}%
\providecommand \@@endlink[0]{}%
\providecommand \url  [0]{\begingroup\@sanitize@url \@url }%
\providecommand \@url [1]{\endgroup\@href {#1}{\urlprefix }}%
\providecommand \urlprefix  [0]{URL }%
\providecommand \Eprint [0]{\href }%
\providecommand \doibase [0]{http://dx.doi.org/}%
\providecommand \selectlanguage [0]{\@gobble}%
\providecommand \bibinfo  [0]{\@secondoftwo}%
\providecommand \bibfield  [0]{\@secondoftwo}%
\providecommand \translation [1]{[#1]}%
\providecommand \BibitemOpen [0]{}%
\providecommand \bibitemStop [0]{}%
\providecommand \bibitemNoStop [0]{.\EOS\space}%
\providecommand \EOS [0]{\spacefactor3000\relax}%
\providecommand \BibitemShut  [1]{\csname bibitem#1\endcsname}%
\let\auto@bib@innerbib\@empty
\bibitem [{\citenamefont {Giorgetta}\ \emph {et~al.}(2013)\citenamefont
  {Giorgetta}, \citenamefont {Swann}, \citenamefont {Sinclair}, \citenamefont
  {Baumann}, \citenamefont {Coddington},\ and\ \citenamefont
  {Newbury}}]{Giorgetta:2013}%
  \BibitemOpen
  \bibfield  {author} {\bibinfo {author} {\bibfnamefont {F.~R.}\ \bibnamefont
  {Giorgetta}}, \bibinfo {author} {\bibfnamefont {W.~C.}\ \bibnamefont
  {Swann}}, \bibinfo {author} {\bibfnamefont {L.~C.}\ \bibnamefont {Sinclair}},
  \bibinfo {author} {\bibfnamefont {E.}~\bibnamefont {Baumann}}, \bibinfo
  {author} {\bibfnamefont {I.}~\bibnamefont {Coddington}}, \ and\ \bibinfo
  {author} {\bibfnamefont {N.~R.}\ \bibnamefont {Newbury}},\ }\href@noop {}
  {\bibfield  {journal} {\bibinfo  {journal} {Nat. Photon.}\ }\textbf {\bibinfo
  {volume} {7}},\ \bibinfo {pages} {434} (\bibinfo {year} {2013})},\ \bibinfo
  {note} {doi:10.1038/nphoton.2013.69}\BibitemShut {NoStop}%
\bibitem [{\citenamefont {Bloom}\ \emph {et~al.}(2014)\citenamefont {Bloom},
  \citenamefont {Nicholson}, \citenamefont {Williams}, \citenamefont
  {Campbell}, \citenamefont {Bishof}, \citenamefont {Zhang}, \citenamefont
  {Zhang}, \citenamefont {Bromley},\ and\ \citenamefont {Ye}}]{Bloom:2014}%
  \BibitemOpen
  \bibfield  {author} {\bibinfo {author} {\bibfnamefont {B.~J.}\ \bibnamefont
  {Bloom}}, \bibinfo {author} {\bibfnamefont {T.~L.}\ \bibnamefont
  {Nicholson}}, \bibinfo {author} {\bibfnamefont {J.~R.}\ \bibnamefont
  {Williams}}, \bibinfo {author} {\bibfnamefont {S.~L.}\ \bibnamefont
  {Campbell}}, \bibinfo {author} {\bibfnamefont {M.}~\bibnamefont {Bishof}},
  \bibinfo {author} {\bibfnamefont {X.}~\bibnamefont {Zhang}}, \bibinfo
  {author} {\bibfnamefont {W.}~\bibnamefont {Zhang}}, \bibinfo {author}
  {\bibfnamefont {S.~L.}\ \bibnamefont {Bromley}}, \ and\ \bibinfo {author}
  {\bibfnamefont {J.}~\bibnamefont {Ye}},\ }\href@noop {} {\bibfield  {journal}
  {\bibinfo  {journal} {Nature}\ }\textbf {\bibinfo {volume} {506}},\ \bibinfo
  {pages} {71} (\bibinfo {year} {2014})},\ \bibinfo {note}
  {doi:10.1038/nature12941}\BibitemShut {NoStop}%
\bibitem [{\citenamefont {Kessler}\ \emph {et~al.}(2012)\citenamefont
  {Kessler}, \citenamefont {Hagemann}, \citenamefont {Grebing}, \citenamefont
  {Legero}, \citenamefont {Sterr}, \citenamefont {Riehle}, \citenamefont
  {Martin}, \citenamefont {Chen},\ and\ \citenamefont {Ye}}]{Kessler:2012}%
  \BibitemOpen
  \bibfield  {author} {\bibinfo {author} {\bibfnamefont {T.}~\bibnamefont
  {Kessler}}, \bibinfo {author} {\bibfnamefont {C.}~\bibnamefont {Hagemann}},
  \bibinfo {author} {\bibfnamefont {C.}~\bibnamefont {Grebing}}, \bibinfo
  {author} {\bibfnamefont {T.}~\bibnamefont {Legero}}, \bibinfo {author}
  {\bibfnamefont {U.}~\bibnamefont {Sterr}}, \bibinfo {author} {\bibfnamefont
  {F.}~\bibnamefont {Riehle}}, \bibinfo {author} {\bibfnamefont {M.~J.}\
  \bibnamefont {Martin}}, \bibinfo {author} {\bibfnamefont {L.}~\bibnamefont
  {Chen}}, \ and\ \bibinfo {author} {\bibfnamefont {J.}~\bibnamefont {Ye}},\
  }\href@noop {} {\bibfield  {journal} {\bibinfo  {journal} {Nat. Photon.}\
  }\textbf {\bibinfo {volume} {6}},\ \bibinfo {pages} {687} (\bibinfo {year}
  {2012})},\ \bibinfo {note} {doi:10.1038/nphoton.2012.217}\BibitemShut
  {NoStop}%
\bibitem [{\citenamefont {Fujieda}\ \emph {et~al.}(2014)\citenamefont
  {Fujieda}, \citenamefont {Piester}, \citenamefont {Gotoh}, \citenamefont
  {Becker}, \citenamefont {Aida},\ and\ \citenamefont {Bauch}}]{Fujieda:2014}%
  \BibitemOpen
  \bibfield  {author} {\bibinfo {author} {\bibfnamefont {M.}~\bibnamefont
  {Fujieda}}, \bibinfo {author} {\bibfnamefont {D.}~\bibnamefont {Piester}},
  \bibinfo {author} {\bibfnamefont {T.}~\bibnamefont {Gotoh}}, \bibinfo
  {author} {\bibfnamefont {J.}~\bibnamefont {Becker}}, \bibinfo {author}
  {\bibfnamefont {M.}~\bibnamefont {Aida}}, \ and\ \bibinfo {author}
  {\bibfnamefont {A.}~\bibnamefont {Bauch}},\ }\href@noop {} {\bibfield
  {journal} {\bibinfo  {journal} {Metrologia}\ }\textbf {\bibinfo {volume}
  {51}},\ \bibinfo {pages} {253} (\bibinfo {year} {2014})},\ \bibinfo {note}
  {doi:10.1088/0026-1394/51/3/253}\BibitemShut {NoStop}%
\bibitem [{\citenamefont {Newbury}\ \emph {et~al.}(2007)\citenamefont
  {Newbury}, \citenamefont {Williams},\ and\ \citenamefont
  {Swann}}]{Newburry:2007}%
  \BibitemOpen
  \bibfield  {author} {\bibinfo {author} {\bibfnamefont {N.~R.}\ \bibnamefont
  {Newbury}}, \bibinfo {author} {\bibfnamefont {P.~A.}\ \bibnamefont
  {Williams}}, \ and\ \bibinfo {author} {\bibfnamefont {W.~C.}\ \bibnamefont
  {Swann}},\ }\href@noop {} {\bibfield  {journal} {\bibinfo  {journal} {Opt.\
  Lett.}\ }\textbf {\bibinfo {volume} {32}},\ \bibinfo {pages} {3056} (\bibinfo
  {year} {2007})},\ \bibinfo {note}
  {http://www.opticsinfobase.org/ol/abstract.cfm?URI=ol-32-21-3056}\BibitemShut
  {NoStop}%
\bibitem [{\citenamefont {Predehl}\ \emph {et~al.}(2012)\citenamefont
  {Predehl}, \citenamefont {Grosche}, \citenamefont {Raupach}, \citenamefont
  {Droste}, \citenamefont {Terra}, \citenamefont {Alnis}, \citenamefont
  {Legero}, \citenamefont {H{\"{a}}nsch}, \citenamefont {Udem}, \citenamefont
  {Holzwarth},\ and\ \citenamefont {Schnatz}}]{Predehl:2012}%
  \BibitemOpen
  \bibfield  {author} {\bibinfo {author} {\bibfnamefont {K.}~\bibnamefont
  {Predehl}}, \bibinfo {author} {\bibfnamefont {G.}~\bibnamefont {Grosche}},
  \bibinfo {author} {\bibfnamefont {S.~M.~F.}\ \bibnamefont {Raupach}},
  \bibinfo {author} {\bibfnamefont {S.}~\bibnamefont {Droste}}, \bibinfo
  {author} {\bibfnamefont {O.}~\bibnamefont {Terra}}, \bibinfo {author}
  {\bibfnamefont {J.}~\bibnamefont {Alnis}}, \bibinfo {author} {\bibfnamefont
  {T.}~\bibnamefont {Legero}}, \bibinfo {author} {\bibfnamefont {T.~W.}\
  \bibnamefont {H{\"{a}}nsch}}, \bibinfo {author} {\bibfnamefont
  {T.}~\bibnamefont {Udem}}, \bibinfo {author} {\bibfnamefont {R.}~\bibnamefont
  {Holzwarth}}, \ and\ \bibinfo {author} {\bibfnamefont {H.}~\bibnamefont
  {Schnatz}},\ }\href@noop {} {\bibfield  {journal} {\bibinfo  {journal}
  {Science}\ }\textbf {\bibinfo {volume} {336}},\ \bibinfo {pages} {441}
  (\bibinfo {year} {2012})},\ \bibinfo {note} {doi:
  10.1126/science.1218442}\BibitemShut {NoStop}%
\bibitem [{\citenamefont {Lopez}\ \emph {et~al.}(2012)\citenamefont {Lopez},
  \citenamefont {Haboucha}, \citenamefont {Chanteau}, \citenamefont
  {Chardonnet}, \citenamefont {Amy-Klein},\ and\ \citenamefont
  {Santarelli}}]{Lopez:2012}%
  \BibitemOpen
  \bibfield  {author} {\bibinfo {author} {\bibfnamefont {O.}~\bibnamefont
  {Lopez}}, \bibinfo {author} {\bibfnamefont {A.}~\bibnamefont {Haboucha}},
  \bibinfo {author} {\bibfnamefont {B.}~\bibnamefont {Chanteau}}, \bibinfo
  {author} {\bibfnamefont {C.}~\bibnamefont {Chardonnet}}, \bibinfo {author}
  {\bibfnamefont {A.}~\bibnamefont {Amy-Klein}}, \ and\ \bibinfo {author}
  {\bibfnamefont {G.}~\bibnamefont {Santarelli}},\ }\href@noop {} {\bibfield
  {journal} {\bibinfo  {journal} {Opt.\ Express}\ }\textbf {\bibinfo {volume}
  {20}},\ \bibinfo {pages} {23518} (\bibinfo {year} {2012})},\ \bibinfo {note}
  {http://dx.doi.org/10.1364/OE.20.023518}\BibitemShut {NoStop}%
\bibitem [{\citenamefont {Droste}\ \emph {et~al.}(2013)\citenamefont {Droste},
  \citenamefont {Ozimek}, \citenamefont {Udem}, \citenamefont {Predehl},
  \citenamefont {H{\"{a}}nsch}, \citenamefont {Schnatz}, \citenamefont
  {Grosche},\ and\ \citenamefont {Holzwarth}}]{Droste:2013}%
  \BibitemOpen
  \bibfield  {author} {\bibinfo {author} {\bibfnamefont {S.}~\bibnamefont
  {Droste}}, \bibinfo {author} {\bibfnamefont {F.}~\bibnamefont {Ozimek}},
  \bibinfo {author} {\bibfnamefont {T.}~\bibnamefont {Udem}}, \bibinfo {author}
  {\bibfnamefont {K.}~\bibnamefont {Predehl}}, \bibinfo {author} {\bibfnamefont
  {T.~W.}\ \bibnamefont {H{\"{a}}nsch}}, \bibinfo {author} {\bibfnamefont
  {H.}~\bibnamefont {Schnatz}}, \bibinfo {author} {\bibfnamefont
  {G.}~\bibnamefont {Grosche}}, \ and\ \bibinfo {author} {\bibfnamefont
  {R.}~\bibnamefont {Holzwarth}},\ }\href@noop {} {\bibfield  {journal}
  {\bibinfo  {journal} {Phys.\ Rev.\ Lett.}\ }\textbf {\bibinfo {volume}
  {111}},\ \bibinfo {pages} {110801} (\bibinfo {year} {2013})}\BibitemShut
  {NoStop}%
\bibitem [{\citenamefont {K\'{e}f\'{e}lian}\ \emph {et~al.}(2009)\citenamefont
  {K\'{e}f\'{e}lian}, \citenamefont {Lopez}, \citenamefont {Jiang},
  \citenamefont {Chardonnet}, \citenamefont {Amy-Klein},\ and\ \citenamefont
  {Santarelli}}]{Kefelian:2009}%
  \BibitemOpen
  \bibfield  {author} {\bibinfo {author} {\bibfnamefont {F.}~\bibnamefont
  {K\'{e}f\'{e}lian}}, \bibinfo {author} {\bibfnamefont {O.}~\bibnamefont
  {Lopez}}, \bibinfo {author} {\bibfnamefont {H.}~\bibnamefont {Jiang}},
  \bibinfo {author} {\bibfnamefont {C.}~\bibnamefont {Chardonnet}}, \bibinfo
  {author} {\bibfnamefont {A.}~\bibnamefont {Amy-Klein}}, \ and\ \bibinfo
  {author} {\bibfnamefont {G.}~\bibnamefont {Santarelli}},\ }\href {\doibase
  10.1364/OL.34.001573} {\bibfield  {journal} {\bibinfo  {journal} {Opt.
  Lett.}\ }\textbf {\bibinfo {volume} {34}},\ \bibinfo {pages} {1573} (\bibinfo
  {year} {2009})}\BibitemShut {NoStop}%
\bibitem [{\citenamefont {Lopez}\ \emph {et~al.}(2013)\citenamefont {Lopez},
  \citenamefont {Kanj}, \citenamefont {Pottie}, \citenamefont {Rovera},
  \citenamefont {Achkar}, \citenamefont {Chardonnet}, \citenamefont
  {Amy-Klein},\ and\ \citenamefont {Santarelli}}]{Lopez:2013}%
  \BibitemOpen
  \bibfield  {author} {\bibinfo {author} {\bibfnamefont {O.}~\bibnamefont
  {Lopez}}, \bibinfo {author} {\bibfnamefont {A.}~\bibnamefont {Kanj}},
  \bibinfo {author} {\bibfnamefont {P.-E.}\ \bibnamefont {Pottie}}, \bibinfo
  {author} {\bibfnamefont {D.}~\bibnamefont {Rovera}}, \bibinfo {author}
  {\bibfnamefont {J.}~\bibnamefont {Achkar}}, \bibinfo {author} {\bibfnamefont
  {C.}~\bibnamefont {Chardonnet}}, \bibinfo {author} {\bibfnamefont
  {A.}~\bibnamefont {Amy-Klein}}, \ and\ \bibinfo {author} {\bibfnamefont
  {G.}~\bibnamefont {Santarelli}},\ }\href {\doibase 10.1007/s00340-012-5241-0}
  {\bibfield  {journal} {\bibinfo  {journal} {Applied Physics B}\ }\textbf
  {\bibinfo {volume} {110}},\ \bibinfo {pages} {3} (\bibinfo {year}
  {2013})}\BibitemShut {NoStop}%
\bibitem [{Han(1989)}]{Hanson:1989}%
  \BibitemOpen
  \href@noop {} {\emph {\bibinfo {title} {Fundamentals of Two-Way Time Transfer
  by Satellite.}}},\ Vol.\ \bibinfo {volume} {Proc. of 43rd Annual Frequency
  Control Symposium}\ (\bibinfo {year} {1989})\BibitemShut {NoStop}%
\bibitem [{\citenamefont {Calosso}\ \emph {et~al.}(2014)\citenamefont
  {Calosso}, \citenamefont {Bertacco}, \citenamefont {Calonico}, \citenamefont
  {Clivati}, \citenamefont {Costanzo}, \citenamefont {Frittelli}, \citenamefont
  {Levi}, \citenamefont {Mura},\ and\ \citenamefont {Godone}}]{Calosso:2014}%
  \BibitemOpen
  \bibfield  {author} {\bibinfo {author} {\bibfnamefont {C.~E.}\ \bibnamefont
  {Calosso}}, \bibinfo {author} {\bibfnamefont {E.}~\bibnamefont {Bertacco}},
  \bibinfo {author} {\bibfnamefont {D.}~\bibnamefont {Calonico}}, \bibinfo
  {author} {\bibfnamefont {C.}~\bibnamefont {Clivati}}, \bibinfo {author}
  {\bibfnamefont {G.~A.}\ \bibnamefont {Costanzo}}, \bibinfo {author}
  {\bibfnamefont {M.}~\bibnamefont {Frittelli}}, \bibinfo {author}
  {\bibfnamefont {F.}~\bibnamefont {Levi}}, \bibinfo {author} {\bibfnamefont
  {A.}~\bibnamefont {Mura}}, \ and\ \bibinfo {author} {\bibfnamefont
  {A.}~\bibnamefont {Godone}},\ }\href@noop {} {\bibfield  {journal} {\bibinfo
  {journal} {Opt.\ Lett.}\ }\textbf {\bibinfo {volume} {39}},\ \bibinfo {pages}
  {1177} (\bibinfo {year} {2014})},\ \bibinfo {note}
  {http://dx.doi.org/10.1364/OL.39.001177}\BibitemShut {NoStop}%
\bibitem [{\citenamefont {Williams}\ \emph {et~al.}(2008)\citenamefont
  {Williams}, \citenamefont {Swann},\ and\ \citenamefont
  {Newbury}}]{Williams:2008}%
  \BibitemOpen
  \bibfield  {author} {\bibinfo {author} {\bibfnamefont {P.~A.}\ \bibnamefont
  {Williams}}, \bibinfo {author} {\bibfnamefont {W.~C.}\ \bibnamefont {Swann}},
  \ and\ \bibinfo {author} {\bibfnamefont {N.~R.}\ \bibnamefont {Newbury}},\
  }\href {\doibase 10.1364/JOSAB.25.001284} {\bibfield  {journal} {\bibinfo
  {journal} {J. Opt. Soc. Am. B}\ }\textbf {\bibinfo {volume} {25}},\ \bibinfo
  {pages} {1284} (\bibinfo {year} {2008})}\BibitemShut {NoStop}%
\bibitem [{\citenamefont {Stefani}\ and\ \citenamefont {{\it et
  al.}}(2014)}]{Stefani:2014}%
  \BibitemOpen
  \bibfield  {author} {\bibinfo {author} {\bibfnamefont {F.}~\bibnamefont
  {Stefani}}\ and\ \bibinfo {author} {\bibnamefont {{\it et al.}}},\
  }\href@noop {} {\  (\bibinfo {year} {2014})},\ \bibinfo {note} {in prep.
  ArXiv:2014/xxxx}\BibitemShut {NoStop}%
\bibitem [{\citenamefont {Bercy}\ \emph {et~al.}(2014)\citenamefont {Bercy},
  \citenamefont {Guellati-Khelifa}, \citenamefont {Stefani}, \citenamefont
  {Santarelli}, \citenamefont {Chardonnet}, \citenamefont {Pottie},
  \citenamefont {Lopez},\ and\ \citenamefont {Amy-Klein}}]{Bercy:2014}%
  \BibitemOpen
  \bibfield  {author} {\bibinfo {author} {\bibfnamefont {A.}~\bibnamefont
  {Bercy}}, \bibinfo {author} {\bibfnamefont {S.}~\bibnamefont
  {Guellati-Khelifa}}, \bibinfo {author} {\bibfnamefont {F.}~\bibnamefont
  {Stefani}}, \bibinfo {author} {\bibfnamefont {G.}~\bibnamefont {Santarelli}},
  \bibinfo {author} {\bibfnamefont {C.}~\bibnamefont {Chardonnet}}, \bibinfo
  {author} {\bibfnamefont {P.-E.}\ \bibnamefont {Pottie}}, \bibinfo {author}
  {\bibfnamefont {O.}~\bibnamefont {Lopez}}, \ and\ \bibinfo {author}
  {\bibfnamefont {A.}~\bibnamefont {Amy-Klein}},\ }\href@noop {} {\bibfield
  {journal} {\bibinfo  {journal} {J.\ Opt.\ Soc.\ Am.\ B}\ }\textbf {\bibinfo
  {volume} {31}},\ \bibinfo {pages} {678} (\bibinfo {year} {2014})},\ \bibinfo
  {note} {http://dx.doi.org/10.1364/JOSAB.31.000678}\BibitemShut {NoStop}%
\bibitem [{\citenamefont {Ma}\ \emph {et~al.}(1994)\citenamefont {Ma},
  \citenamefont {Jungner}, \citenamefont {Ye},\ and\ \citenamefont
  {Hall}}]{Ma:1994}%
  \BibitemOpen
  \bibfield  {author} {\bibinfo {author} {\bibfnamefont {L.~S.}\ \bibnamefont
  {Ma}}, \bibinfo {author} {\bibfnamefont {P.}~\bibnamefont {Jungner}},
  \bibinfo {author} {\bibfnamefont {J.}~\bibnamefont {Ye}}, \ and\ \bibinfo
  {author} {\bibfnamefont {J.~L.}\ \bibnamefont {Hall}},\ }\href@noop {}
  {\bibfield  {journal} {\bibinfo  {journal} {Opt.\ Lett.}\ }\textbf {\bibinfo
  {volume} {19}},\ \bibinfo {pages} {1777} (\bibinfo {year} {1994})},\ \bibinfo
  {note}
  {http://www.opticsinfobase.org/ol/abstract.cfm?URI=ol-19-21-1777}\BibitemShut
  {NoStop}%
\bibitem [{\citenamefont {Jiang}\ \emph {et~al.}(2008)\citenamefont {Jiang},
  \citenamefont {K\'{e}f\'{e}lian}, \citenamefont {Crane}, \citenamefont
  {Lopez}, \citenamefont {Lours}, \citenamefont {Millo}, \citenamefont
  {Holleville}, \citenamefont {Lemonde}, \citenamefont {Chardonnet},
  \citenamefont {Amy-Klein},\ and\ \citenamefont {Santarelli}}]{Jiang:2008}%
  \BibitemOpen
  \bibfield  {author} {\bibinfo {author} {\bibfnamefont {H.}~\bibnamefont
  {Jiang}}, \bibinfo {author} {\bibfnamefont {F.}~\bibnamefont
  {K\'{e}f\'{e}lian}}, \bibinfo {author} {\bibfnamefont {S.}~\bibnamefont
  {Crane}}, \bibinfo {author} {\bibfnamefont {O.}~\bibnamefont {Lopez}},
  \bibinfo {author} {\bibfnamefont {M.}~\bibnamefont {Lours}}, \bibinfo
  {author} {\bibfnamefont {J.}~\bibnamefont {Millo}}, \bibinfo {author}
  {\bibfnamefont {D.}~\bibnamefont {Holleville}}, \bibinfo {author}
  {\bibfnamefont {P.}~\bibnamefont {Lemonde}}, \bibinfo {author} {\bibfnamefont
  {C.}~\bibnamefont {Chardonnet}}, \bibinfo {author} {\bibfnamefont
  {A.}~\bibnamefont {Amy-Klein}}, \ and\ \bibinfo {author} {\bibfnamefont
  {G.}~\bibnamefont {Santarelli}},\ }\href@noop {} {\bibfield  {journal}
  {\bibinfo  {journal} {J.\ Opt.\ Soc.\ Am.\ B}\ }\textbf {\bibinfo {volume}
  {25}},\ \bibinfo {pages} {2029} (\bibinfo {year} {2008})}\BibitemShut
  {NoStop}%
\bibitem [{\citenamefont {Dawkins}\ \emph {et~al.}(2007)\citenamefont
  {Dawkins}, \citenamefont {McFerran},\ and\ \citenamefont
  {Luiten}}]{McFerran:2007}%
  \BibitemOpen
  \bibfield  {author} {\bibinfo {author} {\bibfnamefont {S.}~\bibnamefont
  {Dawkins}}, \bibinfo {author} {\bibfnamefont {J.}~\bibnamefont {McFerran}}, \
  and\ \bibinfo {author} {\bibfnamefont {A.}~\bibnamefont {Luiten}},\
  }\href@noop {} {\bibfield  {journal} {\bibinfo  {journal} {IEEE Trans.
  Ultrason. Ferroelectr. Freq. Control}\ }\textbf {\bibinfo {volume} {54}},\
  \bibinfo {pages} {918} (\bibinfo {year} {2007})}\BibitemShut {NoStop}%
\bibitem [{Ebe(2011)}]{Ebenhag:2011}%
  \BibitemOpen
  \href@noop {} {\emph {\bibinfo {title} {Time Transfer between UTC(SP) and
  UTC(MIKE) Using Frame Detection in Fiber-Optical communication Networks.}}},\
  Vol.\ \bibinfo {volume} {PTTI'11, Long Beach, Ca, 2011-11-14}\ (\bibinfo
  {year} {2011})\BibitemShut {NoStop}%
\bibitem [{\citenamefont {Lee}\ \emph {et~al.}(2010)\citenamefont {Lee},
  \citenamefont {Yu}, \citenamefont {Park},\ and\ \citenamefont
  {Mun}}]{Lee:2010}%
  \BibitemOpen
  \bibfield  {author} {\bibinfo {author} {\bibfnamefont {W.-K.}\ \bibnamefont
  {Lee}}, \bibinfo {author} {\bibfnamefont {D.-H.}\ \bibnamefont {Yu}},
  \bibinfo {author} {\bibfnamefont {C.~Y.}\ \bibnamefont {Park}}, \ and\
  \bibinfo {author} {\bibfnamefont {J.}~\bibnamefont {Mun}},\ }\href {\doibase
  doi:10.1088/0026-1394/47/1/004} {\bibfield  {journal} {\bibinfo  {journal}
  {Metrologia}\ }\textbf {\bibinfo {volume} {47}},\ \bibinfo {pages} {1394}
  (\bibinfo {year} {2010})}\BibitemShut {NoStop}%
\end{thebibliography}%
\end{document}